# A simple contagion process describes spreading of traffic jams in urban networks

Meead Saberi[a,*], Mudabber Ashfaq[a], Homayoun Hamedmoghadam[b], Seyed Amir Hosseini[c], Ziyuan Gu[a], Sajjad Shafiei[d], Divya J. Nair[a], Vinayak Dixit[a], Lauren Gardner[e], S. Travis Waller[a], Marta González[f] a Research Centre for Integrated Transport Innovation (rCITI), School of Civil and Environmental Engineering, University of New South Wales (UNSW), Sydney, NSW 2032, Australia
b Institute of Transport Studies, Civil Engineering Department, Monash University, Melbourne, VIC 3800, Australia
c School of of Electrical and Computer Engineering, K.N. Toosi University of Technology, Tehran, Iran
d Data61, CSIRO, Sydney, Australia
e Department of Civil Engineering, Johns Hopkins University, Baltimore, MD 21218, USA
f College of Environmental Design, University of California, Berkeley, CA 94720, USA

**ABSTRACT**

The spread of traffic jams in urban networks has long been viewed as a complex spatio-temporal phenomenon that often requires computationally intensive microscopic models for analysis purposes. In this study, we present a framework to describe the dynamics of congestion propagation and dissipation of traffic in cities using a simple contagion process, inspired by those used to model infectious disease spread in a population. We introduce two novel macroscopic characteristics of network traffic, namely congestion propagation rate $\beta$ and congestion dissipation rate $\mu$. We describe the dynamics of congestion propagation and dissipation using these new parameters, $\beta$, and $\mu$, embedded within a system of ordinary differential equations, analogous to the well-known Susceptible-Infected-Recovered (SIR) model. The proposed contagion-based dynamics are verified through an empirical multi-city analysis, and can be used to monitor, predict and control the fraction of congested links in the network over time.

## INTRODUCTION

Traffic jams in cities propagate over time and space. Existing approaches to model city traffic often rely on microscopic models with high computational burden as well as excessive parameterization required for calibration [1-3]. Further, the lack of available transport infrastructure data in many countries, especially those that are developing, poses a challenge for traffic modelers. However, the fast-paced development and deployment of mobile sensors offers the opportunity to generate continuous and novel spatial location data, which further enables the estimation of road traffic conditions in real-time.

Numerous studies have recently begun exploring different macroscopic approaches to model the spread of traffic jams in cities [4-9], including through the lens of percolation theory [10, 11]. However, characterizing and modeling congestion propagation and dissipation as a network spreading phenomenon has never been explored. In this work we propose traffic congestion in urban networks can be characterized using a simple contagion process, similar to the well-known Susceptible-Infected-Recovered (SIR) model used to model infectious disease spread in a population, wherein traffic spreads and recovers throughout the network over time (See Fig. 1).

Urban traffic often exhibits high spatial correlation in which links adjacent to a congested link are more likely to become congested. Additionally, the strong temporal correlation of urban congestion is known to be driven by the time-dependent profile of travel demand. The spread of congestion at the link level is well theorized and understood with queuing and kinematic wave theories [12-15]. However, our understanding of congestion propagation dynamics at the network level is still incomplete. Queue spillbacks in networks are shown to be sensitive to link capacities [16] which could remain stable in both under- and over-saturated conditions. Congestion also exhibits fragmentation during recovery [17] leading to greater spatial heterogeneity and thus, results in a drop of network production [18-20]. Propagation and recovery of gridlocks can be characterized by the number of congested links or the length of congestion in the network [18] with propagation occurring often at a much higher rate than dissipation in which the ratio of road supply to travel demand could explain percentage of time lost in congestion [19, 20].

Unlike individual link traffic shockwaves in a two-dimensional time-space diagram, which are categorized as forward or backward moving, network traffic jams evolve in multi directions over space. Therefore, we propose that a network's propagation and recovery can be characterized by two average rates, namely the congestion propagation rate $\beta$ and a congestion recovery rate $\mu$, which together reflect the number of congested links in the network over time. These two macroscopic characteristics are critical in modeling congestion propagation and dissipation as a simple contagion process [21].

Despite the complex human behavior-driven nature of traffic, we demonstrate that urban network traffic congestion follows a surprisingly similar spreading pattern as in other systems, including the spread of infectious disease in a population or diffusion of ideas in a social network, and can be described using a similar parsimonious theoretical network framework. Specifically, we model the spread of congestion in urban networks by adapting a classical epidemic model to include a propagation and recovery mechanism dependent on time-varying travel demand and consistent with fundamentals of network traffic flow theory. We illustrate the model to be a robust and predictive analytical model, and validate the framework using empirical and simulation-based numerical experiments.

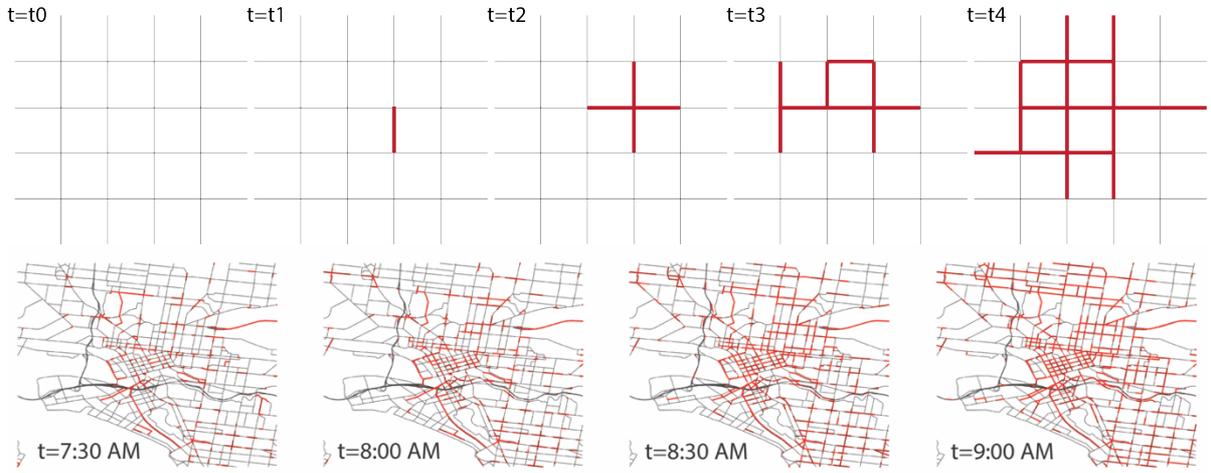

**Fig. 1** Schematic illustration of congestion spreading over time and space in a hypothetical grid network (top row) and simulated road network of Melbourne, Australia (bottom row).

**RESULTS**

We use empirical data from Google that contains estimated time-dependent traffic speeds on every link in the road network across six different cities in the world, namely Chicago, London, Paris, Sydney, Melbourne, and Montreal (*Methods*). We also use simulated data from a calibrated mesoscopic dynamic traffic assignment model of Melbourne (*Methods*) [1]. Using both empirical data and simulation results, we demonstrate that the proposed modeling framework can successfully describe the dynamics of congestion propagation and dissipation in urban networks. In the proposed network-theoretic framework, nodes represent the intersections (controlled and uncontrolled) and links represent the physical roads between any two intersections. See *Methods* and *Supplementary Information* for details.

**Identifying congested links.** For each link in the network, we have the time-dependent speed $v_i(t)$ and the speed limit $v_i^l$. Alternatively, $v_i^l$ can also represent the maximum speed on the link or the 95th percentile of the link speed as in [11]. To reveal how congestion propagates and dissipates in a network, we define

$$r_i(t) = \frac{v_i(t)}{v_i^{max}} \qquad (1)$$

where $r_i(t)$ is the ratio of link speed $v_i(t)$ over link speed limit $v_i^{max}$. We then classify each link as in either a congested $s_i = 1$ or free flow $s_i = 0$ state using a threshold $\rho$ as below

$$s_i(t) = \begin{cases} 1, & r_i(t) < \rho \\ 0, & r_i(t) \geq \rho \end{cases} \quad (2)$$

where $\rho$ represents different congestion levels. We then generate a network using the identified congested links for any given $\rho$ and $t$. Fig. 2 illustrates the identified congested network for various values of $\rho$ at a given $t$ using data from the simulation-based dynamic traffic assignment model of Melbourne. The size of the congested network grows as $\rho$ increases. Alternatively, one can also construct the congested network using traffic density measurements or any other classification method.

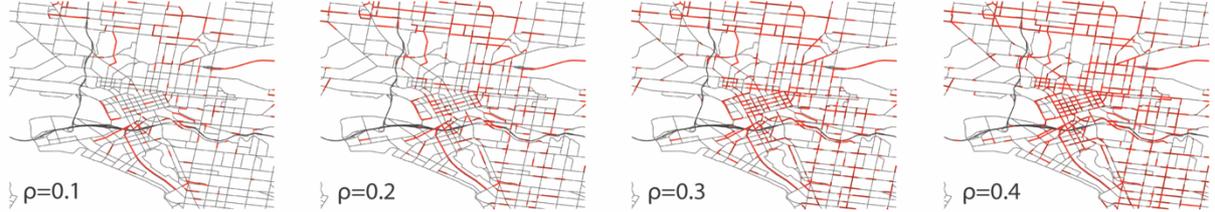

Fig. 2 Spatial distribution of congestion in the network at a given time $t = 8:30$ AM when $\rho = \{0.1, 0.2, 0.3, 0.4\}$.

**Contagion dynamics of network traffic jams.** Consider a network with $N$ directed links. Let $E$ be the set of links, $j$ ($j \in E$) of length $l_j$ in which congestion forms at an individual link $j$ and propagates throughout the network. $F(t)$ represents the number of links that are at free flow regime at time $t$ and $C(t)$ represents the number of congested links at time $t$. At time $t = 0$ every link in the network is at free flow regime, $F(0) = N$ and no link is congested $C(0) = 0$. If a typical directed link in the network is on average connected to $k - 1$ other links at its upstream where $k$ is the average node degree in the network and the rate of congestion propagating to an upstream free flow link in unit time is $\beta$, then assuming a homogenous mixing, the probability of a congested link being connected to and at the downstream of a free flow link is $F(t)/N$. Homogenous mixing refers to an assumption that each link in the network has the same probability of coming into contact with a congested link. Thus, the number of free flow links that are connected to a congested link in a unit time can be expressed as $(k - 1)F(t)/N$. Evidently, in a large-scale network where $N$ is large, the probability of a congested link being connected to an upstream free flow link is close to zero. Although the assumption of homogenous mixing is simplistic, it makes the analysis tractable while shown to be robust and predictive at the macroscopic scale. See *supplementary information* for the extension of the framework when the homogenous mixing assumption is relaxed.

Below we describe the dynamics of congestion propagation with a system of ordinary differential equations (ODE), which is analogous to the well-known SIR model:

$$\frac{dc(t)}{dt} = -\mu c(t) + \beta(k-1)c(t)\big(1 - r(t) - c(t)\big) \quad (3)$$

$$\frac{dr(t)}{dt} = \mu c(t) \quad (4)$$

$$\frac{df(t)}{dt} = -\beta(k-1)c(t)(1 - r(t) - c(t)) \quad (5)$$

where $c(t)$ represents the fraction of congested links in the network, $f(t)$ is the fraction of free flow links, and $r(t)$ is the fraction of recovered links. Eq. 3 describes the rate in which $c(t)$ changes over time given a propagation rate $\beta$ and recovery rate $\mu$ considering that a fraction of congested links will eventually recover as demand for travel diminishes. Eq. 4 expresses the rate in which the fraction of congested links recover given the recovery rate $\mu$. Eq. 5 represents how the fraction of free flow links $f(t)$ in the network changes over time given $c(t)$ and $r(t)$. See Fig. 3. Note that $c(t) + r(t) + f(t) = 1$ where $f(t)$ represents links that have remained free flow from $t = 0$. The formulated model simultaneously describes the dynamics of congestion propagation as well as congestion dissipation in a network, given estimated parameters $\beta$ and $\mu$ which are dependent on a definite time-dependent travel demand profile as in real-world networks. Analogous representations to the well-known susceptible-infected (SI) and susceptible-infected-susceptible (SIS) model of epidemics can also be formulated to describe the spread of traffic in a network. See *Supplementary Information*.

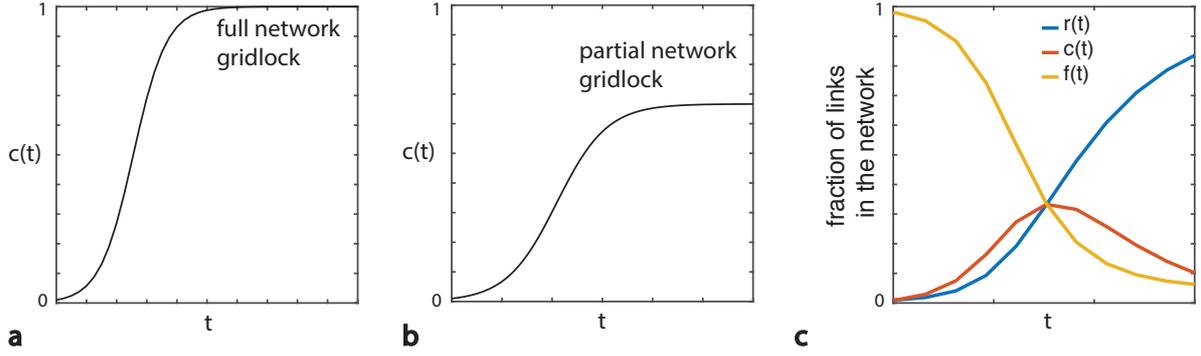

**Fi 3.** Two-state model of congestion propagation in a network: fraction of congested links $c(t)$ vs. time resulting in (a) full network gridlock analogous to the susceptible-infected (SI) model and (b) partial network gridlock analogous to the susceptible-infected-susceptible (SIS) model. (c) Three-state model of congestion propagation and dissipation in a network, analogous to the susceptible-infected-removed (SIR) model subject to a time-varying loading-unloading demand profile.

**Empirical evidence.** We apply the proposed contagion-based model to empirical data collected from six different large metropolitan cities around the world. See Fig. 4. We explore changes in the fraction of congested links $c(t)$ in the networks and the identified congested links. $c(t)$ is calculated as the number of congested links $C(t)$ over the total number of links in the network $N$. The proposed dynamics in Eq. 3-5 are fit to traffic data to estimate the congestion propagation rate $\beta$ and the recovery rate $\mu$ using a pattern search algorithm (*Methods*) for different values of $\rho$ as in Eq. 2.

The applied simple contagion process is illustrated to successfully describes the congestion spreading patterns in different cities at a macroscopic level. While the selected cities have significantly different topology and travel demand patterns, their estimated $\beta/\mu$ ratios are almost the same for $\rho = 0.2$ and slightly vary for $\rho = 0.3$. For larger $\rho$ values, the observed difference between the estimated $\beta/\mu$ ratios grows, which is mainly due to the loose definition of congestion when $\rho$ is large.

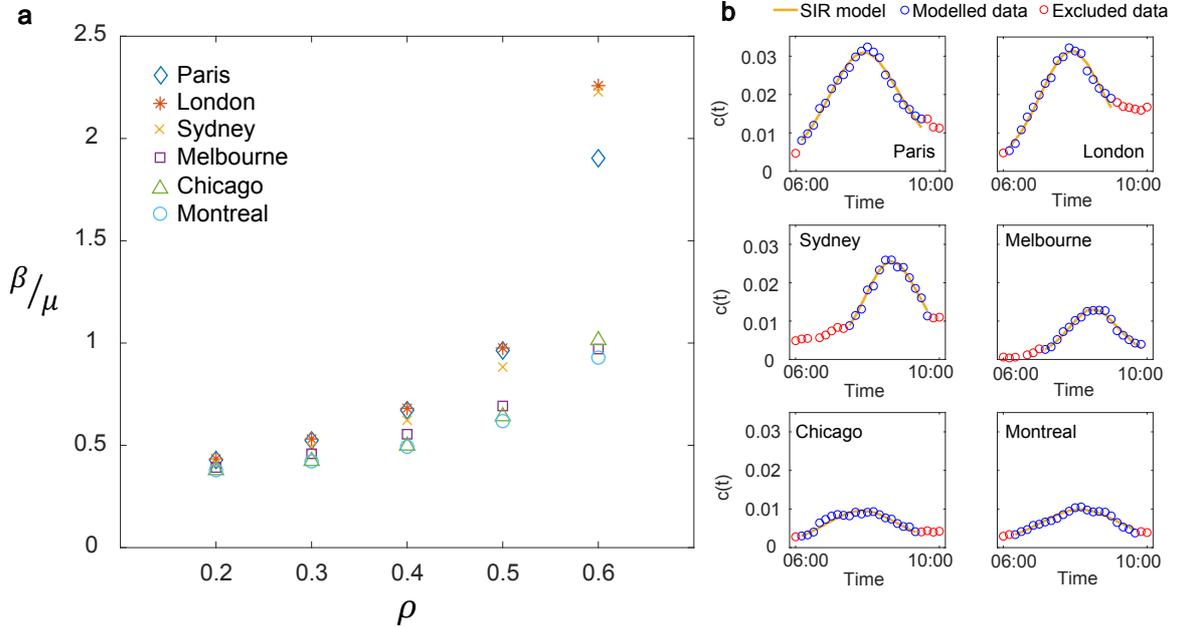

**Fig. 4** Empirical evidence on the congestion propagation and dissipation dynamics from six different cities: (a) relationship between effective congestion spreading rate $\beta/\mu$ and $\rho$. For smaller values of $\rho$, there is a surprising consistency in the observed dynamics across different cities despite their differences in network structure, demand and traffic patterns; (b) proportion of congested links in the network over time $c(t)$ for each city from 6:00 to 10:00 AM when $\rho = 0.2$. Note that the $y$ axis has a fixed range [0, 0.035] and the subset of data within the time frame are utilized to fit the SIR model.

**Revealing the underlying dynamics with simulation.** We now explore changes in the fraction of congested links $c(t)$ in the network given the traffic data obtained from the simulation-based dynamic traffic model of Melbourne and the identified congested links. See *Supplementary Information* for a comparison between empirical and simulation-based data from the Melbourne network. Fig. 5 shows the evolution of $c(t)$ over time when $\rho = \{0.1, 0.3, 0.5, 0.7, 0.9\}$ using simulated data with a calibrated travel demand profile for the morning peak period 6-10 AM followed by a 4-hour recovery period with zero demand. The simulated congestion propagation and dissipation patterns follow the commonly observed spreading patterns in epidemics in which the propagation follows an initial exponential growth regime followed by a sum of multiple exponential processes during recovery. See Fig. 5(a). Note that smaller values of $\rho$ are more appropriate to be used as they better reflect congestion formation compared to larger values of $\rho$. For example, when $\rho = 0.1$ is being used, the fraction of congested links in the network is almost zero for the first hour of the simulation as congestion hasn't yet formed in any link across the network. However, when $\rho = 0.9$ is used, near 15% of the links in the network are already considered congested at the beginning of the simulation.

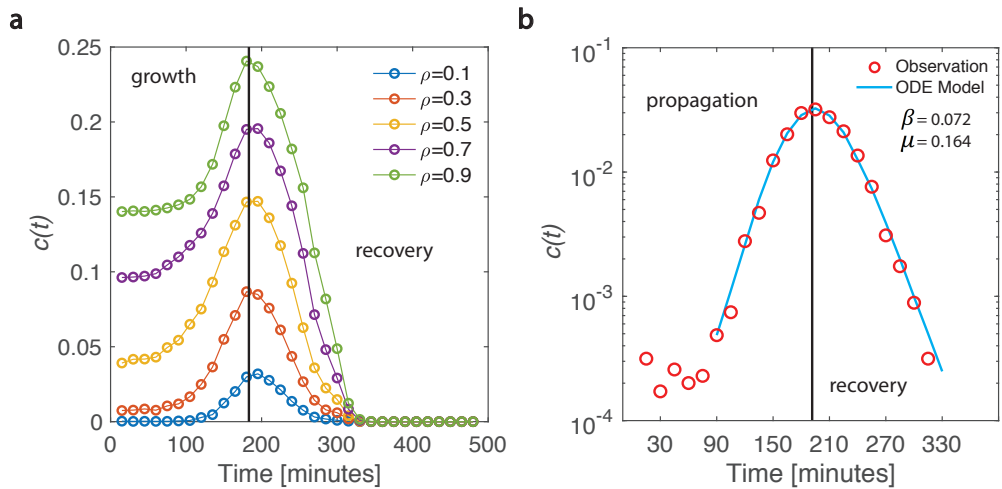

**Fig. 5** Propagation and dissipation of congestion in the Melbourne network subject to realistic and calibrated 4-hr demand followed by a 4-hr recovery period: (a) evolution of fraction of congested links in the network over time for various $\rho$. Data is obtained from the simulation DTA model; (b) fitted modeled dynamics of the system with estimated parameters $\beta = 0.072$ and $\mu = 0.164$ when $\rho = 0.1$. Y axis is scaled logarithmically

To better reveal the congestion spreading patterns, we also conduct a simulation with 1-hour of peak demand loading followed by several hours of recovery with zero demand. See Fig. 6(a) in which congestion propagation in the network can be approximated by an initial exponential growth followed by an exponential decay. Fig. 6(b) illustrates the changes in the estimated parameters $\beta$ and $\mu$ and the ratio of $\beta$ over $\mu$ for a range of $\rho$ values. It is counter-intuitive that when $\rho$ increases, the rates of congestion propagation and dissipation both decrease exponentially. However, the ratio of $\beta$ over $\mu$ increases when $\rho$ increases. In fact, propagation and dissipation rates here must be interpreted relatively. Therefore, their individual values may not provide an absolute indication of the extent that congestion is propagating or dissipating. Instead, it is the ratio of $\beta$ over $\mu$ that has a physical meaning. This is analogous to queuing theory in which the size of a queue depends on the difference between the arrival and departure curves rather than the individual arrival and departure rates. Here, the ratio of $\beta$ over $\mu$ can also be seen as a representation of the rate in which the network shockwave evolves. In other words, $\beta(k-1)/\mu$ represents the average number of newly congested links, each already congested link potentially creates. In epidemics modeling, this is often represented by $R_0$ and called the "basic reproduction number". The higher is $R_0$, the faster congestion spreads throughout the network. If $R_0 \leq 1$, congestion will not spread in the network and remains a non-persistent local phenomenon. Also, $R_0$ follows a linear relationship with $\rho$, as shown in Fig. 6(c), in which when $\rho \to 0$, $R_0 = 1$.

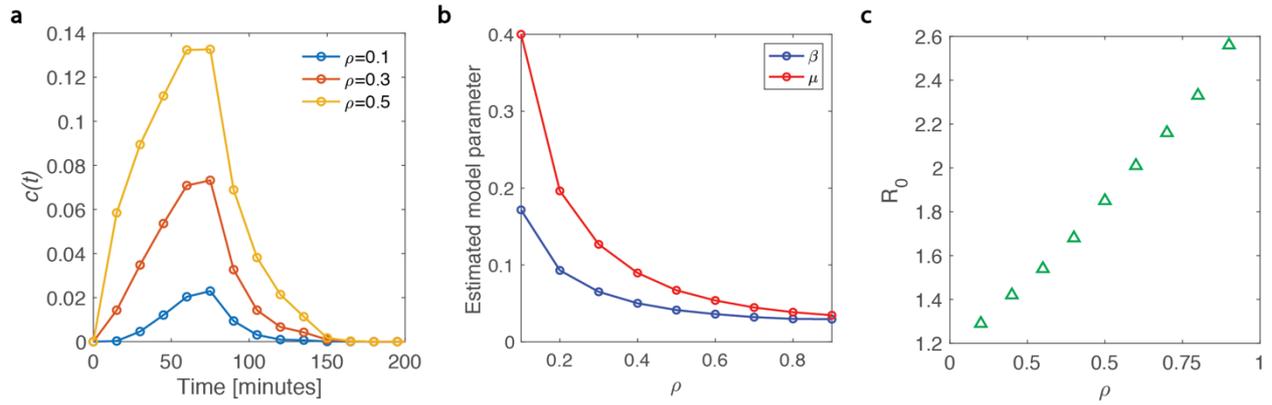

**Fig. 6** Propagation and dissipation of congestion in the Melbourne network subject to 1-hr peak demand followed by a recovery period: (a) evolution of fraction of congested links in the network over time for various $\rho$, (b) estimated model parameters $\beta$ and $\mu$ for various $\rho$, (c) Sensitivity of $R_0$ to $\rho$. Here, $R_0 > 1$ suggesting that congestion is spreading thro

In order to further demonstrate the spreading properties of congestion at the network level, we conduct sensitivity analysis and model evaluation using simulated traffic data. Given a fixed $\rho$, we individually consider each congested link and count the number of upstream congested links at different points in time. We refer to this measurement as the size of the congested upstream cluster. A congested upstream cluster associated with the link $i$, connecting its source node to its target node, includes all the nodes having at least one directed path to the target node traversing the link $i$. For each time step $t$, we generate a null model by randomly drawing $\rho$ of each link from the same $\rho$ distribution as in the simulated data at that time step. To measure how far congestion on a link propagates to the upstream links, we compare the size of the congested upstream cluster associated with each link over time for both the simulated network and the null model. See Fig. 7. Results demonstrate that the spatial distribution of link $\rho$s for the simulated network is significantly different from the null model. The size of the congested upstream clusters in the simulated network changes over time, from low values at $t = 0$, gradually increasing up to approximately $t = 200$ where the cluster size reaches the maximum for a number of links and then decreasing towards zero until around $t = 400$ when the network becomes empty. However, for the null model the size of the clusters is frequently fluctuating over time with large variations around $t = 200$ and $t = 300$ and the maximum cluster size only reaches half of what is observed in the simulation. Consistent with physics of traffic flow and kinematic wave theory, this verifies the hypothesis that congestion at the link level spreads to its upstream links. We also compare the distribution of congested upstream cluster size in the simulation network to that of the null model at a given time $t = 180$ as an example. See Fig. 7(c) for $\rho = 0.5$ and Fig 7(f) for $\rho = 0.7$. The observed difference between the distributions of cluster size in the simulated data versus the null case confirms that congestion follows a non-random spatial spreading pattern.

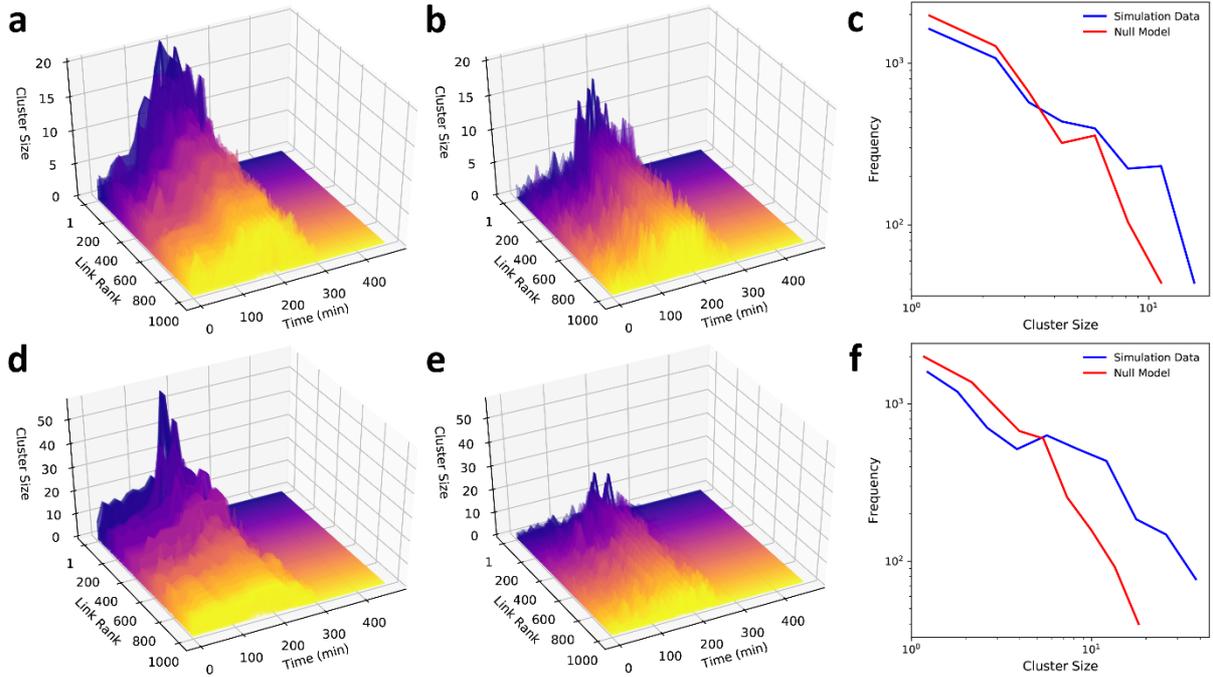

**Fig. 7** Characteristics of the congested upstream cluster associated with network links: (a) and (d) show the change over time in the size of the congested clusters associated with links in the simulation network, while (b) and (e) depict the same results for the null model generated independently for each time step by drawing random $\rho$s for each link from the same distribution as in the simulated data; links are sorted by the maximum size of their congested upstream cluster and the result only includes the top one thousand links. The distribution of congested upstream cluster size at $t$=180 is depicted in (c) and (f) for the Melbourne traffic simulation model and its null model counterpart. Top and bottom panels correspond to $\rho = 0.5$ and $\rho = 0.7$ respectively.

**Connection with travel demand.** Here, we examine the impact of changing demand on the propagation and dissipation dynamics of congestion. For the same network with 1-hour of loading followed by 4-hour recovery, we have conducted multiple simulation runs with 75%, 100%, 125%, 150%, 175%, and 200% demand. We then solve the proposed ODE model and estimate the model parameters $\beta$ and $\mu$ with the global pattern search as described previously. When demand increases, the fraction of congested links also increases for the same network and as a result, complete recovery of congestion takes longer while the start of the recovery phase remains the same. This is reflected in the reduction of $\mu$ in response to the increase in demand for any given $\rho$. See Fig. 8(a) for the evolution of $c(t)$ over time for various demand levels. Counter-intuitively, increasing demand also results in reduction of $\beta$ for a given $\rho$. See Fig. 8(b) and 8(c). This should not be interpreted as a slower propagation of congestion. In fact, $\beta$ is not independent of $\mu$. The two macroscopic characteristics vary interdependently, so $c(t)$ reaches its peak value at $t = 75$ min according to the demand profile. However, the ratio of $\beta$ over $\mu$ increases when demand increases for a given $\rho$ with a clear indication that the size of the network shockwave will grow larger and it takes longer to recover. Interestingly, $\beta/\mu$ has an exponential response to demand $\beta/\mu \propto e^{\alpha.demand}$ in which a relative change in demand results in an exponential change in $\beta/\mu$. See Fig. 8(d) and 8(e). Results suggest that there is also a three-dimensional relationship between the ratio of $\beta$ over $\mu$, $\rho$, and demand level as illustrated in Fig. 8(f), in which for smaller values of $\rho$, the relationship between $\beta/\mu$ and demand is almost linear.

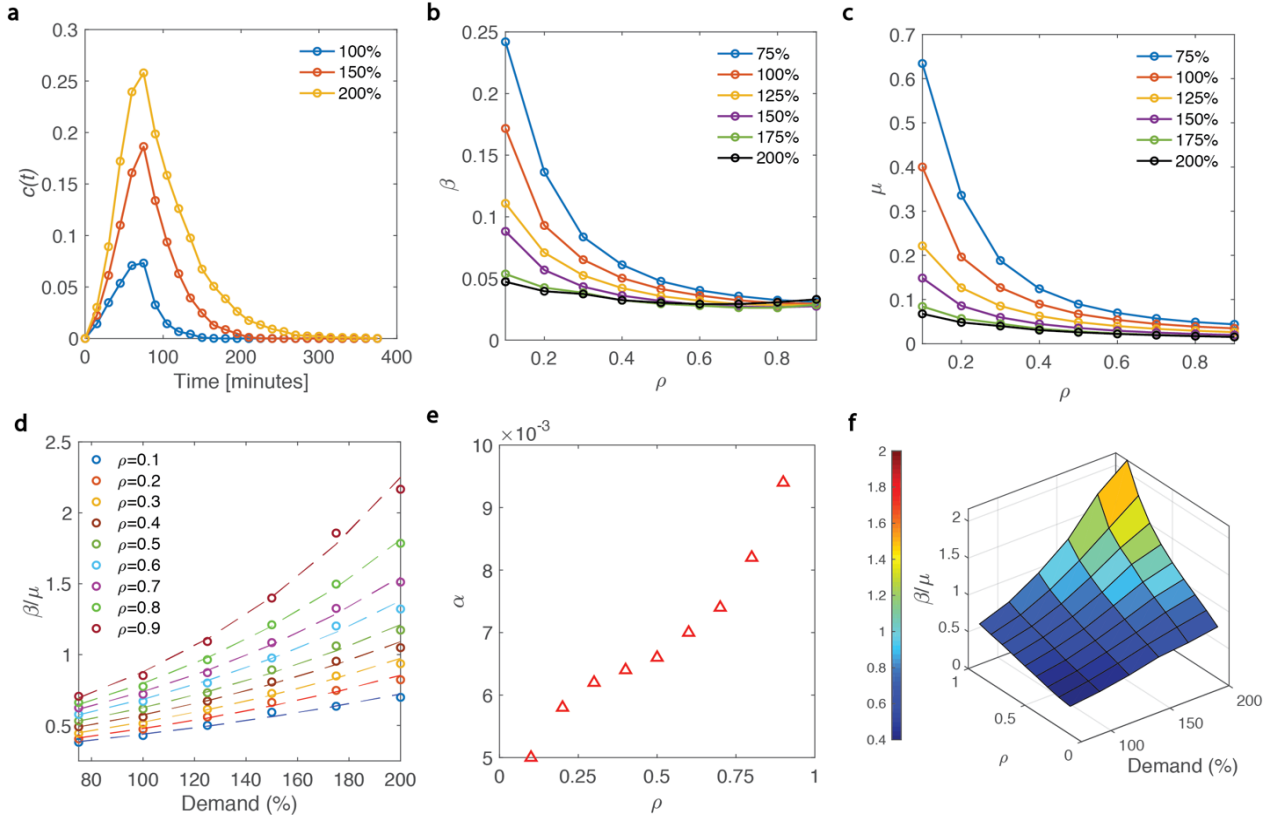

**Fig. 8** Propagation and dissipation of congestion in the Melbourne network subject to 1-hr demand followed by a recovery period: (a) evolution of fraction of congested links in the network over time for various $\rho$; (b) and (c) estimated model parameters $\beta$ and $\mu$ for various $\rho$ and demand levels; (d) Ratio of estimated $\beta$ over $\mu$ as a function of demand for various $\rho$ values. Dashed lines represent the exponential fit to the data; (e) Sensitivity of $\alpha$ to $\rho$. Here $\alpha$ represents the exponent of the exponential curves fitted in (d). (f) three-dimensional illustration of the relationship between the ratio of $\beta$ over $\mu$, $\rho$ and demand.

## DISCUSSION

This study has shown that propagation and dissipation of traffic jams in cities can be described by a simple contagion process which is formulated as a system of ordinary differential equations, analogous to the well-known SIR model, dependent on two macroscopic network traffic flow characteristics, namely congestion propagation and congestion recovery rate. The paper has shown both empirically and simulation-based that congestion propagation is indeed a spreading phenomenon. The extent in which congestion builds up in a network and how fast it recovers are shown to be dependent on the ratio of the propagation over recovery rates, similar to arrival and departure rates in a queuing system. Using data from a simulation-based dynamic traffic assignment model of Melbourne, we showed that time-dependent travel demand profile has, as expected, an impact on dynamics of congestion propagation and dissipation in a network. In fact, the ratio of $\beta$ over $\mu$ increases when demand increases for any given $\rho$ with a clear indication that the size of the traffic jam will grow larger and it takes longer to recover following an exponential law in respond to travel demand. Moreover, empirical data of six different cities are analyzed to verify the validity of the proposed contagion-based model. We observed that different cities surprisingly exhibit very similar $\beta/\mu$ ratio for small $\rho$ values, indicating that at the macroscopic level, they tend to have consistent congestion propagation dynamics. A limitation of this study is that in the classical SIR model when an infected individual recovers, he/she will not be infected again (or removed). This clearly does not apply directly to traffic networks in which a link may recover and become congested again after a short period. However, the assumption may still hold if the temporal aggregation of analysis is large enough to prevent links to go under multiple cycles of congestion and recovery in a short period. Given the unimodality of travel demand profile during a peak period (morning or afternoon), it may not be unreasonable to assume, at a macroscopic level, that when a link recovers from congestion, it will not become congested again until the next peak period. Nevertheless, relaxation of this assumption and introduction of other traffic states considering multiple cycles of loading and recovery should be an item of research priority in future.

## METHODS

**Traffic simulation model.** Simulation of the Melbourne network is conducted in AIMSUN. The model is a mesoscopic simulation-based dynamic traffic assignment which has been calibrated and validated for the morning peak period 6-10 AM. A large number of input (demand and supply) parameters need to be calibrated before the simulation outcomes are used. Details of the calibration and validation process can be found in [9] and supplementary information.

**Google traffic data.** Google traffic data on 27/06/2018 are acquired and used which consist of speeds in the unit of km/h for each link in the network with a unique identifier. The time frame for congestion modeling varies from city to city, which is determined based on the profile of the fraction of congested links, so as to get a fitted model. Nevertheless, the time remains the same as in our simulation (i.e. 6-10 AM). Links with missing data for the entire day are discarded in our analysis, but the number of such links is negligible and does not affect the results.

Google traffic data record the speeds of road sections at continuous time intervals. The data were estimated using floating GPS points of mobile users once they used Google services such as Google Maps. The sampling interval of the data is 10 seconds which provides a good temporal resolution for our analysis. Cities that are selected include Melbourne, Sydney, London, Paris, Chicago and Montreal, which all have a complex urban transport network with a dense population, thereby ensuring the availability of GPS data in immense quantity (note that the reliability of estimations is directly proportional to the amount of data). The calculation of $v_i^{\max}$ takes into account the maximum speed observed on road segment $i$ in a cycle of one day (i.e. 24 hours). Using this measure along with $v_i(t)$, the ratio $r_i(t)$ is calculated for the entire network. To be consistent, only the morning peak of each city is considered in our analysis which all lies in the time period between 6:00 and 10:00 AM, although being subject to some differences from city to city.

**Parameter estimation with global pattern search.** To estimate the parameters of the proposed model, we have applied a "global pattern search" algorithm as a derivative-free global optimization method in which the modeled curves for $c(t)$ are fit to the simulated data with the objective function to minimize the root mean squared error (RMSE). The pattern search algorithm falls under the general category of global optimization methods in which an initial mesh is first specified in the solution space given an initial guessed solution. The algorithm computes the objective function value at each mesh point until it finds a point whose value is smaller than the objective function value at the initial solution point. The algorithm then updates the mesh size and re-computes the objective function value at each mesh point. This will iteratively continue until one of the stopping criteria is met such as the mesh size getting smaller than a mesh tolerance threshold or if the maximum number of iterations is reached. For details about the pattern search algorithm, see [22]. Also see Table S1 in the Supplementary Information for the estimated model parameters and associated RMSE for different values of $\rho$.

**Data availability.** For contractual and privacy reasons, we cannot make the empirical data from Google available. However, all data from the simulation model that are needed to replicate the findings reported in the paper and supplementary information are available at https://github.com/meeadsaberi/trafficspreading. The simulation-based dynamic traffic assignment model is also available as an open access model which can be downloaded via https://github.com/meeadsaberi/dynamel.

## REFERENCES

1. Shafiei, S., Gu, Z., Saberi, M., 2018. Calibration and validation of a simulation-based dynamic traffic assignment model for a large-scale congested network. Simulation Modelling Practice and Theory 86, 169-186.
2. Olmos, L.E., Çolak, S., Shafiei, S., Saberi, M., Gonzalez, M.C., 2018. Macroscopic dynamics and the collapse of urban traffic. Proceedings of the National Academy of Science 115 (50), 12654-12661
3. Jiang, S., et al. (2016) The TimeGeo modeling framework for urban mobility without travel surveys. Proc. Natl Acad. Sci. (USA) 113(37):E5370–E5378.
4. Geroliminis, N., Daganzo, C., 2008. Existence of urban-scale macroscopic fundamental diagrams: Some experimental findings. Transportation Research Part B 42(9), 759-770.
5. Gayah, V., Daganzo, C., 2011. Clockwise hysteresis loops in the Macroscopic Fundamental Diagram: An effect on network instability. Transportation Research Part B 45(4), 643-655.
6. Saberi, M., Mahmassani, H., 2012. Exploring properties of network-wide flow-density relations in a freeway network. Transportation Research Record 2315, 153-163.
7. Saberi, M., Mahmassani, H., 2013. Hysteresis and Capacity Drop Phenomena in Freeway Networks: Empirical Characterization and Interpretation. Transportation Research Record 2391, 44-55.
8. Ji, Y., Geroliminis, N., 2012. On the spatial partitioning of urban transportation networks. Transportation Research Part B: Methodological 46(10), 1639-1656.

9. Saeedmanesh, M., Geroliminis, N., 2017. Dynamic clustering and propagation of congestion in heterogeneously congested urban traffic networks. Transportation Research Part B: Methodological 105, 193-211.
10. Zeng, G., Li, D., Guo, S., Gao, L., Gao, Z., Stanley, H.E., Havlin, S., 2019. Switch between critical percolation modes in city traffic dynamics. Proceedings of the National Academy of Science 116 (1), 23-28.
11. Li, D., Fu, B., Wang, Y., Lu, G., Berezin, Y., Stanley, H.E., Havlin, S., 2015. Percolation transition in dynamical traffic network with evolving critical bottlenecks. Proceedings of the National Academy of Sciences of the United States of America 112(3), 669–672.
12. Lighthill M. J. and Whitham J. B., 1955. On kinematic waves. I: Flow movement in long rivers; II: A theory of traffic flow on long crowded roads. Proc. Royal Sot. A, 229, 281-345.
13. Richards P. I., 1956. Shockwaves on the highway. Opns. Res. 4,42-51.
14. Newell G. F., 1993. A simplified theory of kinematic waves. 1: general theory; II: Queuing at freeway bottlenecks; III: Multi-destination flows. Transportation Research 27B, 281-314.
15. Daganzo, C., 1994. The Cell Transmission model: A dynamic representation of highway traffic consistent with the hydrodynamic theory. Transportation Research 28B, 269-287.
16. Daganzo, C., 1998. Queue spillovers in transportation networks with a route choice. Transportation Science 32 (1), 3–11.
17. Daganzo, C., 2011. On the macroscopic stability of freeway traffic. Transportation Research Part B: Methodological 45, 782–788.
18. Mahmassani, H., Saberi, M., Zockaie, A., 2013. Urban network gridlock: Theory, characteristics, and dynamics. Transportation Research Part C: Emerging Technologies 36, 480-497.
19. Çolak, S., Lima, A., González, M., 2016. Understanding congested travel in urban areas. Nature Communications volume 7, Article number: 10793.
20. Olmos, L.E., Çolak, S., Shafiei, S., Saberi, M., González, M. (2018) Macroscopic dynamics and the collapse of urban traffic. Proceedings of the National Academy of Sciences of the United States of America 11 (50), 12654-12661.
21. Brockmann, D., Helbing, D., 2013. The Hidden Geometry of Complex, Network-Driven Contagion Phenomena. Science 342(6164), 1337-1342.
22. Audet, C., Dennis, J.E., 2003. Analysis of Generalized Pattern Searches. SIAM Journal on Optimization 13(3), 889-903.

**ACKNOWLEDGEMENTS**

Authors are thankful for the fruitful discussions with Dr. David Rey at University of New South Wales and Dr. Mohsen Ramezani from University of Sydney.

**AUTHOR CONTRIBUTIONS**

M.S., M.G., and L.G. designed the study. M.S., M.A., H.H., and S.A.H. implemented the method. M.A., S.S. and Z.G. conducted the simulations. D.J.N. and V.D. supported with empirical data acquisition and analysis. M.S., M.A., H.H., S.A.H., Z.G., L.G., S.T.W. and M.G. analyzed the results and wrote the manuscript.